\documentclass[aps,pra,twocolumn,shownopacs]{revtex4}
\usepackage{graphicx}
\usepackage{hyperref}
\usepackage{amssymb,amsmath}
\usepackage{epstopdf}
\usepackage{ulem}

\begin{document}

\title{Enhancing electromagnetically-induced transparency\\in a multilevel broadened medium}

\author{M. Scherman, O.S. Mishina, P. Lombardi, E. Giacobino, and J. Laurat}

\affiliation{Laboratoire Kastler Brossel,
Universit\'{e} Pierre et Marie Curie, Ecole Normale Sup\'{e}rieure,
CNRS, Case 74, 4 place Jussieu, 75252 Paris Cedex 05, France}

\date{\today}

\begin{abstract}
Electromagnetically-induced
transparency
has become an important tool to control the optical properties of
dense media. 
However, in a broad class of systems, the interplay between inhomogeneous broadening and the
existence of several excited levels may lead to a vanishing
transparency. Here, by identifying the underlying
physical mechanisms resulting in this effect, we show 
that transparency
can be strongly enhanced. We thereby demonstrate a 5-fold enhancement in a
room-temperature vapor of alkali-metal atoms via a
specific shaping of the atomic velocity distribution. 
\end{abstract}

\maketitle

\section{Introduction}
Among the capabilities enabled by electromagnetically induced transparency (EIT)  \cite{Boller,Fleischhauer2005} or related
phenomena are high precision magnetometery,
lasing without inversion, and slowing
\cite{Hau1999} and stopping of light pulses
\cite{Liu2001,Phillips2001}. These possibilities opened new
avenues for optical information storage and quantum information
processing. Recent experiments based on dynamic EIT have
demonstrated the reversible mapping of single-photons or qubits
\cite{Matsukevich2005,Eisaman2005,Choi2008} and of quantum
continuous variables \cite{Honda2008,Appel2008,Cviklinski2008}.
These seminal demonstrations spurred intense experimental and
theoretical efforts to improve the efficiency of such processes
and extend them to new enabling photonic technologies.

An important effort concerns indeed the modeling of the EIT
process in the non-ideal case. The EIT configuration is usually
modeled by a generic $\Lambda$-type three-level system,
 most relevant for quantum memory schemes \cite{Fleischhauer2005}: two atomic
ground states are connected to an excited state via two optical
fields, a probe and a control field. However, in many optically
dense media, the relevant energy structure is more complex and can
strongly modify the EIT features. A typical case is the use of
ensembles of alkali-metal atoms in which experiments have been
most performed \cite{Lvovsky2009,Hammerer2010}. The hyperfine
interaction in the excited state introduces several levels, which
can simultaneously participate in the coherent interaction. The
deviation from the $\Lambda$-type approximation can be very
significant when the inhomogeneous broadening is comparable with
the separation between these excited levels, such as for example
in the D$_2$-line of atomic cesium at room temperature. The
observed transparency is generally lower than predicted
\cite{Cviklinski2008,Akulshin1998} and can eventually disappear
for large broadening \cite{Li2009}. The broadening also leads to a narrowing of the transparency window \cite{Ye2002,Hossain2009,Figueroa2006}. Various
numerical analysis have investigated particular regimes, such as
double-$\Lambda$ system \cite{Cerboneschi1996} or off-resonant
Raman transition in a broadened medium \cite{Sheremet2010}.
However, to date no full study of EIT in inhomogeneously broadened
medium with multiple excited levels has been performed.

In the present paper, we report measurements that provide a detailed
picture of EIT in a Doppler broadened medium. In agreement with
the general model recently developed in Ref. \cite{Oxy2011}, we
evidence the process leading to a reduced transparency and we
experimentally demonstrate how to mitigate this effect. Our
observations are made possible by identifying atoms from specific
velocity classes that absorb the light in the process and by then
reshaping accordingly the atomic velocity distribution. This
procedure enables to recover a significant transparency.

\section{Effect of inhomogeneous broadening on
the EIT features}
 In a $\Lambda$-type system, the
susceptibility of the medium is strongly modified when a driving
field is applied to one of the transition. In the absence of
broadening, the transmission of a probe field exhibits two symmetrical absorption peaks as a function of
its detuning from the resonance, defining a
transparency window at resonance. In an inhomogeneously broadened
medium, the absorption spectrum of the various atoms differs from
this description as they are involved in off-resonant
processes, drastically modifying the susceptibility and the EIT spectrum.

\begin{figure*}[htpb!]
\includegraphics[width=1.5\columnwidth]{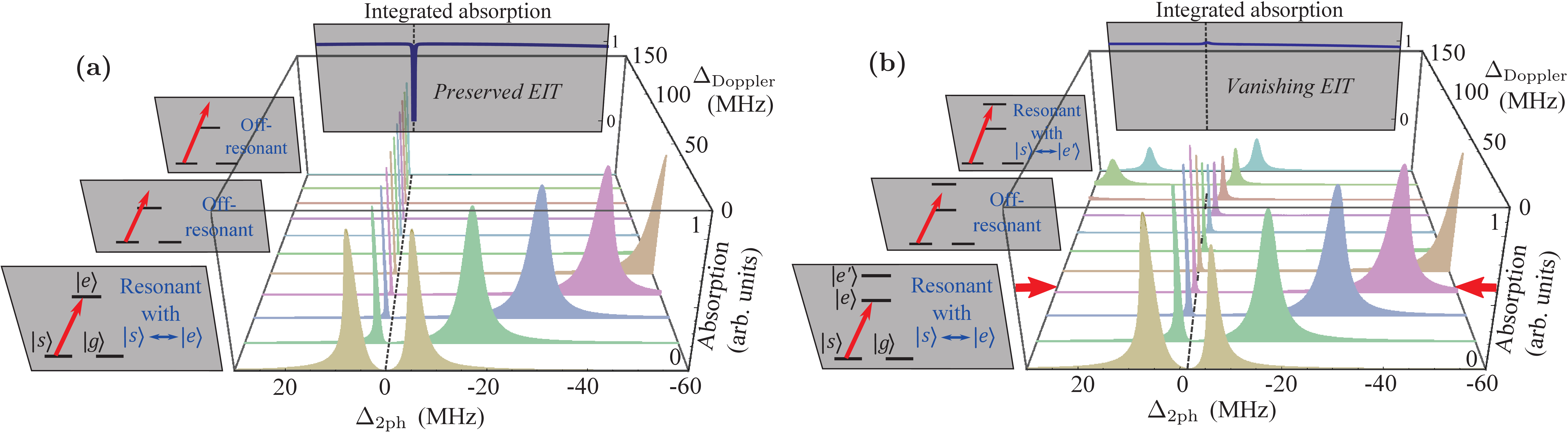}
\caption{EIT features in a $\Lambda$ and in a multi-level system. The calculated probe absorptions are displayed for atoms
with different velocities  given by the Doppler detunings
$\Delta_{Doppler}$, as a function of the detuning $\Delta_{2ph}$. Panel (a)
corresponds to a $\Lambda$-scheme and shows a preserved transparency.
Panel (b) takes into account an additional excited state, leading
to a vanishing transparency. The dotted line indicates the
center of the EIT window for atoms with $\Delta_{Doppler}$=0. The
integrated absorption is obtained for a Gaussian velocity
distribution of 160 MHz half-width (thermal distribution for Cesium at 300 K). The control Rabi
frequency is $\Omega=2.3\gamma $, where $\gamma $ is the
natural linewidth.} \label{schema_3D}
\end{figure*}

Specifically, in a Doppler broadened medium, the atoms are distributed over a wide range of velocity classes. For an atom
moving with velocity $\textbf{v}$ and copropagating control and
probe fields, the two laser frequencies are Doppler shifted in the
atom rest frame by approximately the same detuning
$2\pi.\Delta_{Doppler}=-\textbf{k}\cdot\textbf{v}$ where
$\textbf{k}$ is the wave-vector of the fields. The two-photon
detuning, defined as
$2\pi.\Delta_{2ph}=\omega_{probe}-\omega_{control}+\omega_{gs}$
where $\omega_{probe}$ and $\omega_{control}$ are the field
frequencies and $\omega_{gs}$ the splitting between the two ground states, does not depend on the velocity either.
Figure \ref{schema_3D}(a) gives the absorption as a function of
the two-photon detuning for different Doppler shifts when the
control field is resonant for atoms at rest. We
have considered the case of velocities opposite to the
laser propagation direction ($\Delta_{Doppler}>0$). While for
atoms at rest one can observe the usual Autler-Townes doublet, the
spectrum is modified when the Doppler shift increases, i.e. the
two absorption peaks are not symmetrical anymore. The peak with
$\Delta_{2ph}<0$ corresponds to the one-photon absorption
resonance and the second peak ($\Delta_{2ph}>0$) corresponds to
the Raman absorption process, and it is Stark shifted from the
zero two-photon detuning. When the Doppler detuning increases,
this Raman peak gets closer to $\Delta_{2ph}=0$ without reaching
this point. As a result, transparency is preserved at the
zero two-photon detuning for all the velocity classes. The
integration over the whole distribution thus preserves the transparency at resonance but
leads to a reduction of the transparency window width.

When one takes into account the presence of other levels,
 additional Stark shifts appear that move the Raman resonance from the
position described in the $\Lambda$ configuration.  Based on the
model developed in \cite{Oxy2011}, Fig. \ref{schema_3D}(b) shows
the case of two excited levels, $|e\rangle$ and $|e'\rangle$. The
strong modification of the susceptibility can be understood as
follows. In this scheme, two velocity classes now see the control
field on resonance but for different levels: $|e\rangle$ for zero
Doppler shift and $|e'\rangle$ for a shift equal to the separation
between the two levels \textbf{$\omega_{e'e}$}. These two classes
exhibit a quasi-symmetrical Autler-Townes doublet centered
close to their respective atomic transition, i.e. on the
zero two-photon detuning shown on Fig. 1(b). For atoms with
intermediate velocities, the Raman absorption undergoes Stark
shifts due to the two excited states and eventually crosses the
transparency window of atoms with zero Doppler shift, as
illustrated in Fig. \ref{schema_3D}(b). While transparency was
always preserved at resonance in the $\Lambda$ model, here there
is no longer any value of the detuning for which atoms are
transparent independently of their velocity. Consequently, if the
broadening is comparable with the hyperfine splitting, the
integration over all the velocity classes can result in a total
disappearance of the EIT, as shown here. We find that in our case the atomic velocity
classes to be removed for optimal EIT recovery correspond to
Doppler detunings 35MHz$\lesssim\Delta_0\lesssim$45MHz (Appendix A). By excluding these specific atoms from the
interaction process, the EIT can be recovered, as we will show.

\begin{figure*}[htpb!]
\includegraphics[width=1.5\columnwidth]{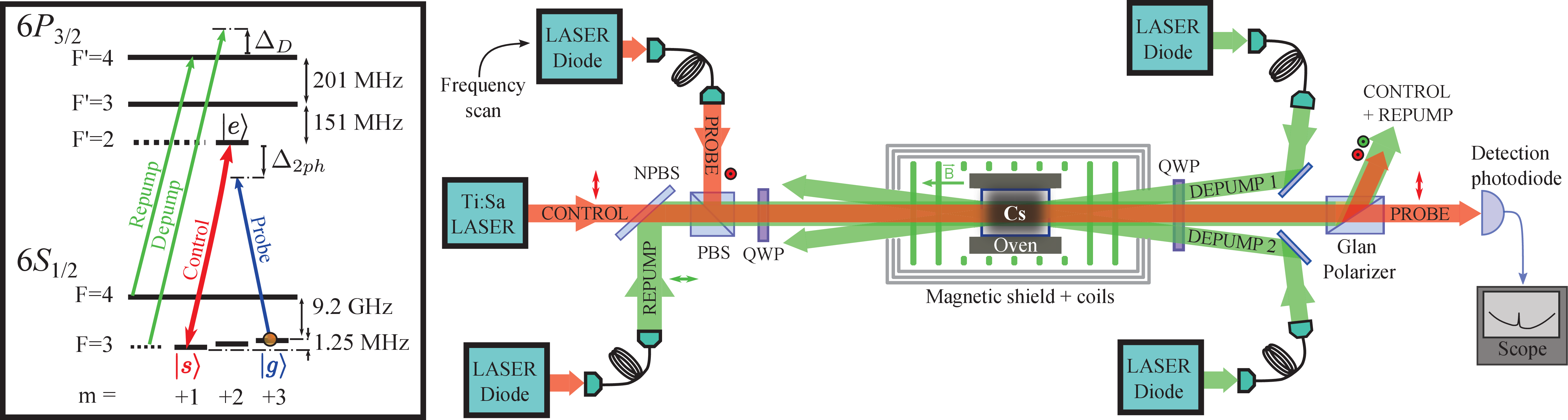}
\caption{The transmission of a cesium vapor is probed by
scanning the frequency of a weak $\sigma^-$-polarized probe field.
The strong $\sigma^+$-polarized control field is kept on resonance
with the $|s\rangle$ to $|e\rangle$ transition. A repumper beam
enables to efficiently prepare the atoms in the $|g\rangle$ ground
state. Two  $\sigma^+$-polarized depumpers beams can be used to
burn holes in the velocity distribution in order to exclude atoms
with specific Doppler shifts from the interaction process. PBS and NPBS : polarizing and
non-polarizing beam splitter, QWP: $\lambda/4$ plate.}
\label{schema_manip}\vspace{-0.4cm}
\end{figure*}

\section{Experimental setup and results}
The experiment is sketched in Fig.
\ref{schema_manip}. The optically dense medium is obtained from a vapor of $^{133}$Cs heated at 35$^{\circ}$C in a
paraffin-coated cylindrical glass cell (3 cm long and 3 cm in
diameter). At this temperature, the Doppler broadening reaches a
half-width equal to 160 MHz. The cell is placed in a longitudinal
magnetic field produced by sets of coils and the system is
enclosed into a magnetic shield. The scheme of the interaction
is given in the inset of Fig. \ref{schema_manip}. The
two ground states $|s\rangle$ and $|g\rangle$ are the two Zeeman states
$|6S_{1/2},F=3,m=1\rangle$ and $|6S_{1/2},F=3,m=3\rangle$ separated by 1.25 MHz. The excited
state $|e\rangle$ is the Zeeman state $|6P_{3/2},F=2,m=2\rangle$. As explained
previously, this simple $\Lambda$ system is strongly influenced by
the other excited states, i.e. $|6P_{3/2},F=3\rangle$ and $|6P_{3/2},F=4\rangle$ which are
respectively 151 MHz and 352 MHz from the $|e\rangle$ state. The control
field is $\sigma^{+}$-polarized and resonant with the transition
$|s\rangle \rightarrow |e\rangle$ for non-moving atoms, while the probe field is
$\sigma^{-}$-polarized and addresses the transition $|g\rangle \rightarrow
|e\rangle$.

The experiment is performed in the continuous-wave regime. The
control field tends to pump all the atoms into the Zeeman sublevel
of maximum $m$ of the F=3 level. However, off-resonant pumping
through the excited state $|6P_{3/2},F=3\rangle$ in the presence of large
Doppler shifts can eventually drive all the atomic population into
the $|F=4\rangle$ dark state. To prevent this depumping, an additional
$\sigma^{+}$ repump field is used on the transition
$|F=4\rangle \rightarrow |F'=4\rangle$. In this way, a significant fraction of the
atoms is maintained in $|6S_{1/2},F=3\rangle$ \cite{Cviklinski2008}.

Control and probe fields are collimated (5mm diameter) and with respective powers of 200 mW and 150 nW. They are
combined on a polarizing beam splitter and then pass through a
$\lambda/4$ to enter the cell with circular polarization.
The probe field is then extracted using a $\lambda/4$ and a
Glan polarizer with high extinction ratio ($10^6$). The repump
beam has a slightly larger beam size and its power is adjusted
during the experiment to keep the optical density constant.

As explained above, the disappearance of the EIT is predicted to
be due to some specific velocity classes. In order to exclude
these atoms from the interaction, we use one or two additional
depumping beams  detuned by $\Delta_D$ from the
$|F=3\rangle \rightarrow |F'=4\rangle$ transition for non-moving atoms. They enable to burn specific holes in the velocity distribution.
For experimental convenience, they are contra-propagative with the
control and probe fields, with an angle around 3$^{\circ}$.
In order to remove atoms that see the probe field frequency
shifted by $\Delta_{0}$, a counterpropagating depumping beam must
have a frequency detuned by $\Delta_D=\Delta_{0}$. The atoms
are then efficiently pumped into the hyperfine ground state F=4
and do not contribute anymore to the interaction.

\begin{figure*}[htpb!]
\includegraphics[width=1.5\columnwidth]{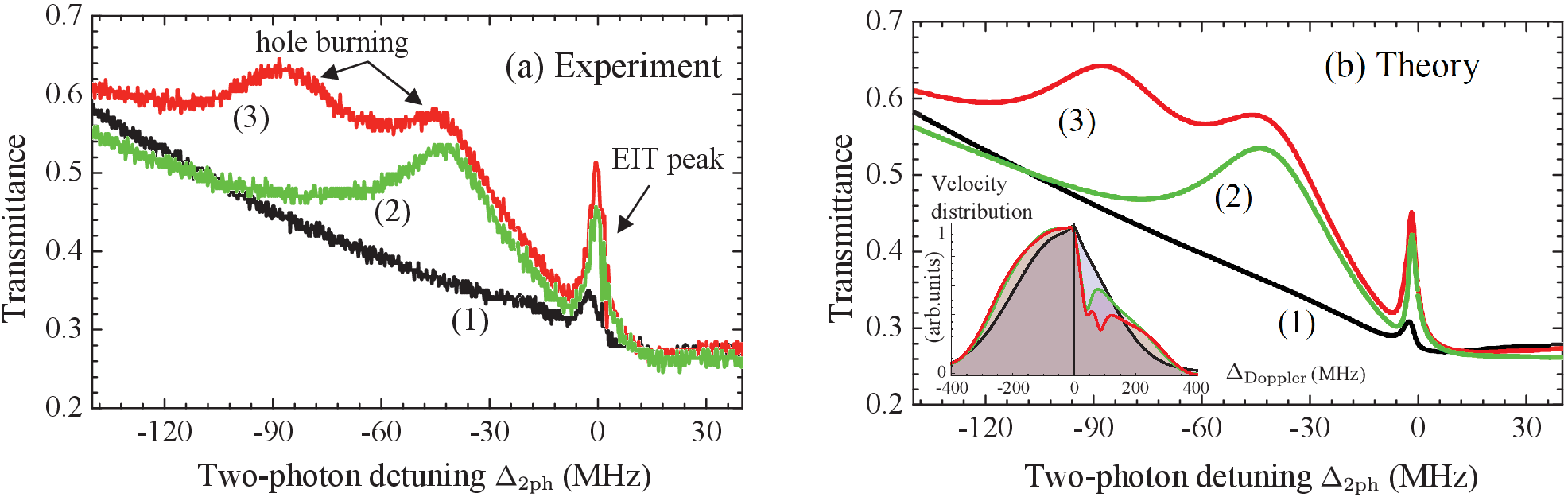}
\caption{Transparency enhancement by reshaping of the velocity distribution. (a) Transmission of the probe field as a function of the detuning $\Delta_{2ph}$. Curve
(1): without depumping beams, Curve (2): one depumping beam with
$\Delta_D$=40 MHz and Curve (3): two depumping beams with
$\Delta_D$=40 MHz and 85 MHz. The transmission is
normalized to the transmission for large detunings. The
experimental powers are 6 mW for the depumpers, 4.5 mW for the
repumper without depump, $5.5$ mW with one depumps and $7.5$ mW
with two depumps (see text). (b) Theoretical predictions
calculated from Ref. \cite{Oxy2011}. $\Omega= 2.3\gamma$,
corresponding to the experimental value, with $\gamma=2\pi\times
5.2$ MHz. We include the ground state decoherence due to the dephasing between control and probe lasers experimentally estimated to be $\gamma_{sg}=0.077\gamma$. This value corresponds to twice the linewidth of these independent lasers. The
inset gives the velocity distribution extracted from the data and
used for the model.} \label{courbe_exp_theo}\vspace{-0.3cm}
\end{figure*}

Transmission spectra are obtained by scanning the probe
detuning, which corresponds to scanning the two-photon detuning
$\Delta_{2ph}$. Figure \ref{courbe_exp_theo}(a) gives the spectrum, with and without reshaping of the atomic
velocity distribution. Curve (1) shows a very weak EIT peak near
$\Delta_{2ph}=0$ as expected in such a broadened medium with
several excited states. We then send a depumping beam.
A broad peak, corresponding to the hole burnt in the velocity
distribution, appears in the transmission spectrum (curve (2)). As can be seen, the EIT peak is significantly enhanced.
The best transparency is obtained for detunings close to the
predicted value of $\Delta_D=$ 40 MHz, confirming that
the EIT is recovered when the detrimental effect of these atoms is
suppressed. The depth of the hole created in the
distribution by the depumping beam at 40 MHz being
saturated for a power of about 6mW, we tested the effect of a
second depumping beam, detuned from the optimal frequency and
corresponding to atoms with $\Delta_D=$85 MHz. As shown by curve
(3), the transparency is slightly increased, since this beam removes some
more atoms which have non-zero absorption in the EIT window.

\section{Discussion and conclusion}
These results are in very good agreement with the transmission
curves calculated from the model developed in Ref. \cite{Oxy2011} and given in Fig. \ref{courbe_exp_theo}(b). In this case, we have
used the velocity distributions extracted from the experimental
data to compute the susceptibility of the atomic ensemble, as
shown in the inset of Fig. \ref{courbe_exp_theo}(b). Even in the
absence of depumping beams, the distribution is non
Gaussian, since it is strongly modified by the control and the
repump beams \cite{Smith,Macquire,Lindvall}. It can be seen that
the effect of one and two depumping beams on the EIT is correctly
reproduced. Because of the distorted velocity distribution, this
model also correctly predicts a small EIT feature in the absence
of depumping, in contrast to the model of Fig. \ref{schema_3D}(b).

Finally, in order to quantify the EIT enhancement, we
introduce  an EIT contrast $C$ defined as the ratio
$C=(t_{max}-t_{min})/(1-t_{min})$ with $t_{max}$ and $t_{min}$
being the probe transmittance at the maximum and on
the side of the EIT peak.  In our experiment, $C$ is increased by
a factor  5, which can be shown to yield a decrease of the group
velocity by a factor of the same order of magnitude. This shows
the potential of our method for efficiently improving
EIT-based processes. We note that the full theoretical model \cite{Oxy2011} predicts an enhancement factor around 8. The main limitation in our experimental demonstration is coming from the contrast of the hole burned in the velocity distribution. A detailed study of the hole burning dynamics is expected to bring further enhancement of the EIT feature \cite{Smith,Macquire} .

In summary, we have reported a detailed experimental
characterization of the EIT properties of a medium with Doppler
broadening and multiple excited levels. It can be shown that some
specific velocity classes are mainly causing the suppression of
transparency. We have proposed a procedure to remove these atoms
from the interaction via a well-designed reshaping of the
atomic velocity distribution. Our observations confirmed the
general mechanism and enabled to demonstrate a strong enhancement
of the transparency. This study may bring new applications and
offer significant improvements in various settings based on EIT or
related effects, including in quantum information science and
metrology. Moreover, beyond the specific medium used here, i.e
alkali-metal atoms at room temperature, our method, which allows
an efficient engineering of the EIT properties of an
inhomogeneously broadened medium, can be extended to various
atom-like physical systems presenting simultaneously large
broadening and multiple levels, e.g. in rare-earth doped crystals,
quantum dots or  nitrogen-vacancy centers in diamonds.

\appendix
\section{Determination of velocity classes leading to absorption}
\noindent The simplified model given in the text enables to identify the atoms which
strongly modify the EIT and to calculate their
Doppler shift $\Delta_{0}$. With $d$ and $d'$ the dipole
elements for $|s\rangle\rightarrow |e\rangle$ and
$|s\rangle\rightarrow |e'\rangle$ transitions and $\omega_{e'e}$
the splitting in the excited state, $\Delta_{0}$ can be
obtained from:
\begin{equation}\label{1}
    \frac{|d|^2}{\Delta_{0}}+
    \frac{|d'|^2}{(\Delta_{0}-\omega_{e'e})}
    =
    -\frac{|d'|^2}{\omega_{e'e}.}\nonumber
\end{equation}
This equation states that the Raman absorption resonance, displaced by the
Stark shifts due to the two excited levels (left hand side) is
centered at the same frequency as the EIT window for atoms with
zero Doppler shift, itself shifted due to level $|e'\rangle$
(right hand side). This equation has two solutions that do
not depend on the control field Rabi frequency. The solution that
introduces the largest absorption is $\Delta_{0}$=48MHz, marked by
arrows in Fig. \ref{1}.

With a full model \cite{Oxy2011} taking into account the
hyperfine structure of atomic cesium, and the variation of
the height of the Raman absorption resonance with Doppler
detuning, we find that in our case the atomic velocity
classes to be removed for optimal EIT recovery correspond to
Doppler detunings 35MHz$\lesssim\Delta_0\lesssim$45MHz, which is consistent with the simplified

\begin{acknowledgements} We thank D. Felinto, J.~Ortalo and S.~Burks for
fruitful discussions. This work is supported by the EC under the
ICT/FET project COMPAS. O.S. Mishina acknowledges the financial
support of the Ile-de-France programme IFRAF. J. Laurat is member of the Institut Universitaire de France.
\end{acknowledgements}

\end{document}